\documentclass[graybox]{svmult}

\usepackage{mathptmx}       % selects Times Roman as basic font
\usepackage{helvet}         % selects Helvetica as sans-serif font
\usepackage{courier}        % selects Courier as typewriter font
\usepackage{type1cm}        % activate if the above 3 fonts are
\usepackage{makeidx}         % allows index generation
\usepackage{graphicx}        % standard LaTeX graphics tool
\usepackage{subfigure}
\usepackage{multicol}        % used for the two-column index
\usepackage[bottom]{footmisc}% places footnotes at page bottom

\makeindex             % used for the subject index
\begin{document}

\title*{A possible GeV-radio correlation for starburst galaxies}
% Use \titlerunning{Short Title} for an abbreviated version of
% your contribution title if the original one is too long
\author{S.~Sch\"oneberg, J.~Becker~Tjus, F.~Schuppan}
% Use \authorrunning{Short Title} for an abbreviated version of
% your contribution title if the original one is too long
\institute{S.~Sch\"oneberg \at Ruhr-Universit\"at Bochum, Fakult\"at f\"ur Physik \& Astronomie, 44780 Bochum, Germany \\
\email{ssc@tp4.rub.de}
\and J.~Becker~Tjus \at Ruhr-Universit\"at Bochum, Fakult\"at f\"ur Physik \& Astronomie, 44780 Bochum, Germany
\and F.~Schuppan \at Ruhr-Universit\"at Bochum, Fakult\"at f\"ur Physik \& Astronomie, 44780 Bochum, Germany}

% Use the package "url.sty" to avoid
% problems with special characters
% used in your e-mail or web address
%
\maketitle

\abstract{For star-forming regions, there is a correlation of radio 
		and FIR-emission established. The radio emission is caused by 
		synchrotron radiation of electrons, while the FIR emission is attributed 
		to HII regions of OB stars and hot dust powered by those stars. 
		Another possible correlation for star-forming regions might exist 
		between the aforementioned radio emission and the gamma ray emission 
		in the GeV regime. 
		The GeV gamma ray emission of star-forming regions is dominated by 
		the decay of neutral pions formed in collisions of cosmic ray (CR) protons 
		accelerated at supernova remnants (SNRs) with ambient hydrogen, while the 
		electrons generating the synchrotron emission are assumed to be 
		accelerated at the same SNRs. Assuming the same spectral shape for 
		the CR proton and electron distribution and a fixed ratio of CR protons 
		to electrons, the proton- and electron-calorimetry of the star-forming 
		regions can be tested. Furthermore, this provides a method to derive the 
		magnetic field strength in the star-forming region complementary to 
		the existing methods. }

%\abstract{Each chapter should be preceded by an abstract (10--15 lines long) that summarizes the content. The abstract will appear \textit{online} at \url{www.SpringerLink.com} and be available with unrestricted access. This allows unregistered users to read the abstract as a teaser for the complete chapter. As a general rule the abstracts will not appear in the printed version of your book unless it is the style of your particular book or that of the series to which your book belongs.\newline\indent
%Please use the 'starred' version of the new Springer \texttt{abstract} command for typesetting the text of the online abstracts (cf. source file of this chapter template \texttt{abstract}) and include them with the source files of your manuscript. Use the plain \texttt{abstract} command if the abstract is also to appear in the printed version of the book.}

\section{Introduction}
\label{sec:1}
\begin{figure}
	\centering
		\includegraphics[scale=2.0]{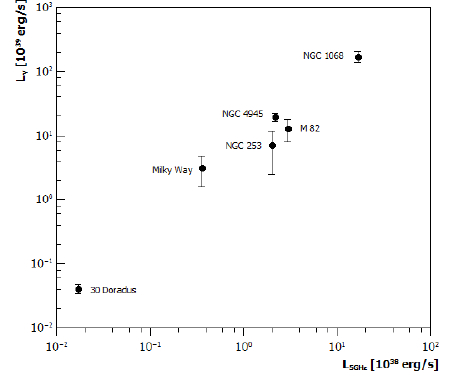}
	\caption{Correlation between radio luminosity at 5 GHz and the
$\gamma$-luminosity.}
	\label{ssc::RG}
\end{figure}

A plot of the total gamma ray luminosity against the luminosity at 5~GHz for 
M82 and NGC 253 (both starburst galaxies), NGC 1068 and NGC 4945 (starburst-
Seyfert composites) and 30 Doradus in the Large Magellanic Cloud and the Milky 
Way center reveals a trend, as Fig.\ \ref{ssc::RG} shows.
To interpret this possible correlation, a theoretical approach is followed:
The emission of the objects dealt with is assumed to be dominated by SNRs inside the objects. The detected radio emission is attributed to synchrotron radiation of electrons accelerated at the SNRs, the GeV gamma rays are caused by the decay of neutral pions formed in collisions of protons accelerated at the SNRs with ambient protons:
\begin{center}
\begin{eqnarray}
 {\mathrm{p}}_{\mathrm{CR}} + {\mathrm{p}} \rightarrow \pi^{+/-/0} + X\\
 \pi^{0} \rightarrow \gamma + \gamma~. \,\,\,\,\,\,\,\,\,\,\,\,\,\,\,\,\,\,\,
\end{eqnarray}
\end{center}
Neutrinos are produced from charged pions, where searches are dominated by a background of atmospheric neutrinos so far \cite{Anatoli2012}. Both protons and electrons are probably accelerated at the same site, so it is assumed that their spectra have the same --or a very similar-- shape, they only differ by a constant factor
\begin{center}
\begin{eqnarray}
\frac{{\mathrm{d}}N}{{\mathrm{d}}E_{\mathrm{p}}} = a_{\mathrm{p}}  \left(\frac{E_{\mathrm{p}}}{E_0}\right)^{-p} , \,\,\,
\frac{{\mathrm{d}}N}{{\mathrm{d}}E_{\mathrm{e}}} = a_{\mathrm{e}} \left(\frac{E_{\mathrm{e}}}{E_0}\right)^{-p} , \,\,\,
0.01 \leq \frac{a_{\mathrm{e}}}{a_{\mathrm{p}}} \leq 1. 
\end{eqnarray}
\label{spec_shape}
\end{center}
To examine the trend in Fig.\ \ref{ssc::RG}, both radiation processes are investigated in more detail in the following sections.

%Use the template \emph{chapter.tex} together with the Springer document class SVMono (monograph-type books) or SVMult (edited books) to style the various elements of your chapter content in the Springer layout.

%Instead of simply listing headings of different levels we recommend to
%let every heading be followed by at least a short passage of text.
%Further on please use the \LaTeX\ automatism for all your
%cross-references and citations. And please note that the first line of
%text that follows a heading is not indented, whereas the first lines of
%all subsequent paragraphs are.

\section{Electron synchrotron radiation $\rightarrow$ Radio emission}
\label{sec:2}
The electron spectral energy distribution is assumed to be a simple power-law:  
\begin{equation}
N(E_{\mathrm{e}}) \,{\mathrm{d}}E_{\mathrm{e}}=a_{\mathrm{e}} \left(\frac{E_{\mathrm{e}}}{m_{\mathrm{e}} c^2}\right)^{-p}\,{\mathrm{d}}E~. 
\end{equation}
If each electron radiates all its energy at a single frequency $\nu$, the emission coefficient $\epsilon_{\nu}$ can be written as:
\begin{center}
\begin{eqnarray}
\Rightarrow \epsilon_\nu = a_{\mathrm{e}} \frac16 \beta^2 m_{\mathrm{e}}^2 c^4 \left(\frac{\sigma_{\mathrm{T}} B}{e} \right)
\left(\frac{\nu}{\nu_{\mathrm{G}}}\right)^\delta~, \,\,\,\,\,\,\,\,\,\,\,\,\,\,\,\,\,\,\,\,\,\,\,\,\,\,\,\,\,\,\,\,\,\,\,\,\,\,\,\,\,\,\,\,\,\,\,\,\,\,\,\,\,\,\,\,\,\,\,\,\,\,\,\,\,\, \\
E = \gamma m_{\mathrm{e}}c^2 ,\,\, \gamma = \left(\frac{\nu}{\nu_{\mathrm{G}}} \right)^{\frac12} ,\,\, \frac{{\mathrm{d}}E}{{\mathrm{d}}\nu} = \frac{\nu^{-\frac12}}{2\nu_{\mathrm{G}}^{\frac12}}m_{\mathrm{e}}c^2 ,\,\, U_B = \frac{B^2}{8\pi} ,\,\, \delta = (1-p)/2~,
\end{eqnarray}
\end{center}
using the expressions above, where $e$ is the elementary charge, $\nu_{\mathrm{G}}$ is the gyrofrequency and $\sigma_{\mathrm{T}}$ is the Thomson cross-section.
In order to obtain the radio luminosity of an object, the emission coefficient is multiplied by the average volume $V$ of an SNR and the number of SNRs, $N_{\mathrm {SNR}}$ in the object:
\begin{eqnarray}
L_\textrm{Radio} = V N_{\mathrm {SNR}} \,\int\limits_{\nu_0}^{\nu_1} \epsilon_\nu(\nu) \,
{\mathrm{d}}\nu \,\,\,\,\,\,\,\,\,\,\,\,\,\,\,\,\,\,\,\,\,\,\,\,\,\,\,\,\,\,\,\,\,\,\,\,\,\,\,\,\,\,\,\,\,\,\,\,\,\,\,\,\,\,\,\,\,\,\,\,\,\,\,\,\,\,\,\,\\
{\mathrm{d}}L_\nu = V N_{\mathrm {SNR}} \, a_{\mathrm{e}} \frac16 \beta^2 m_{\mathrm{e}}^2 c^4 \left(\frac{\sigma_{\mathrm{T}} B}{e}
\right)\left(\frac{\nu}{\nu_{\mathrm{G}}}\right)^\delta \, {\mathrm{d}} \nu~.
\end{eqnarray}
By expressing the emission coefficient in terms of energy instead of frequency, the synchrotron flux can be written as:
\begin{eqnarray}
E^2 \frac{{\mathrm{d}}N}{{\mathrm{d}}E} = \frac{E}{h} \epsilon(E) = \frac{E}{h} A_{\mathrm{e}} \frac16 \beta^2 m_{\mathrm{e}}^2 c^4
\left(\frac{\sigma_{\mathrm{T}} B}{e} \right)\left(\frac{E}{h\nu_{\mathrm{G}}}\right)^\delta~,
\end{eqnarray}
where $A_{\mathrm{e}} = V N_{\mathrm {SNR}} \, a_{\mathrm{e}}$.

\section{GeV gamma rays from pp-interaction}
\label{sec:3}
Following \cite{kelner_pp2006}, the differential proton flux in each object is derived from gamma ray observations under the assumption that these gamma rays are dominantly formed by pion decay from proton-proton interactions. The spectral shape of the primary proton flux is the same as that of the primary electrons (see eq.\ \ref{spec_shape}) and of the gamma rays from pion decay in the energy range of $300~
\textrm{MeV}~<~E_\gamma~<~E_{\mathrm{p}}^\textrm{max}/10$.
The differential gamma ray luminosity is calculated as:
\begin{equation}
E_\gamma^2 \frac{{\mathrm{d}}N_\gamma}{{\mathrm{d}}E_\gamma} = V N_{\mathrm {SNR}} a_{\mathrm{p}} n_{\mathrm{H}} c E_\gamma^2 \int\limits_{E_{\min}}^{\infty} \sigma_{\mathrm{inel}} (E_{\mathrm{p}}) \left(\frac{E_{\mathrm{p}}}{E_0} \right)^{-p}
F_{\gamma} \left(\frac{E_\gamma}{E_{\mathrm{p}}}, \, E_{\mathrm{p}} \right) \, \frac{{\mathrm{d}}E_{\mathrm{p}}}{E_{\mathrm{p}}}~.
\label{ssc::gammalum}
\end{equation}
The calculation is done following \cite{kelner_pp2006}, where the $\delta$-functional approach is used for energies below 100~GeV and the analytical approximation given there is used for higher energies.

% Always give a unique label
% and use \ref{<label>} for cross-references
% and \cite{<label>} for bibliographic references
% use \sectionmark{}
% to alter or adjust the section heading in the running head

%Instead of simply listing headings of different levels we recommend to
%let every heading be followed by at least a short passage of text.
%Further on please use the \LaTeX\ automatism for all your
%cross-references and citations.

%Please note that the first line of text that follows a heading is not indented, whereas the first lines of all subsequent paragraphs are.

%Use the standard \verb|equation| environment to typeset your equations, e.g.
%
%\begin{equation}
%a \times b = c\;,
%\end{equation}
%
%however, for multiline equations we recommend to use the \verb|eqnarray| environment\footnote{In physics texts please activate the class option \texttt{vecphys} to depict your vectors in \textbf{\itshape boldface-italic} type - as is customary for a wide range of physical subjects}.
%\begin{eqnarray}
%a \times b = c \nonumber\\
%\vec{a} \cdot \vec{b}=\vec{c}
%\label{eq:01}
%\end{eqnarray}

\section{Comparison and conclusions}
\label{sec:4}

\begin{figure}[htbp]
\begin{center}
\includegraphics[width = \textwidth]{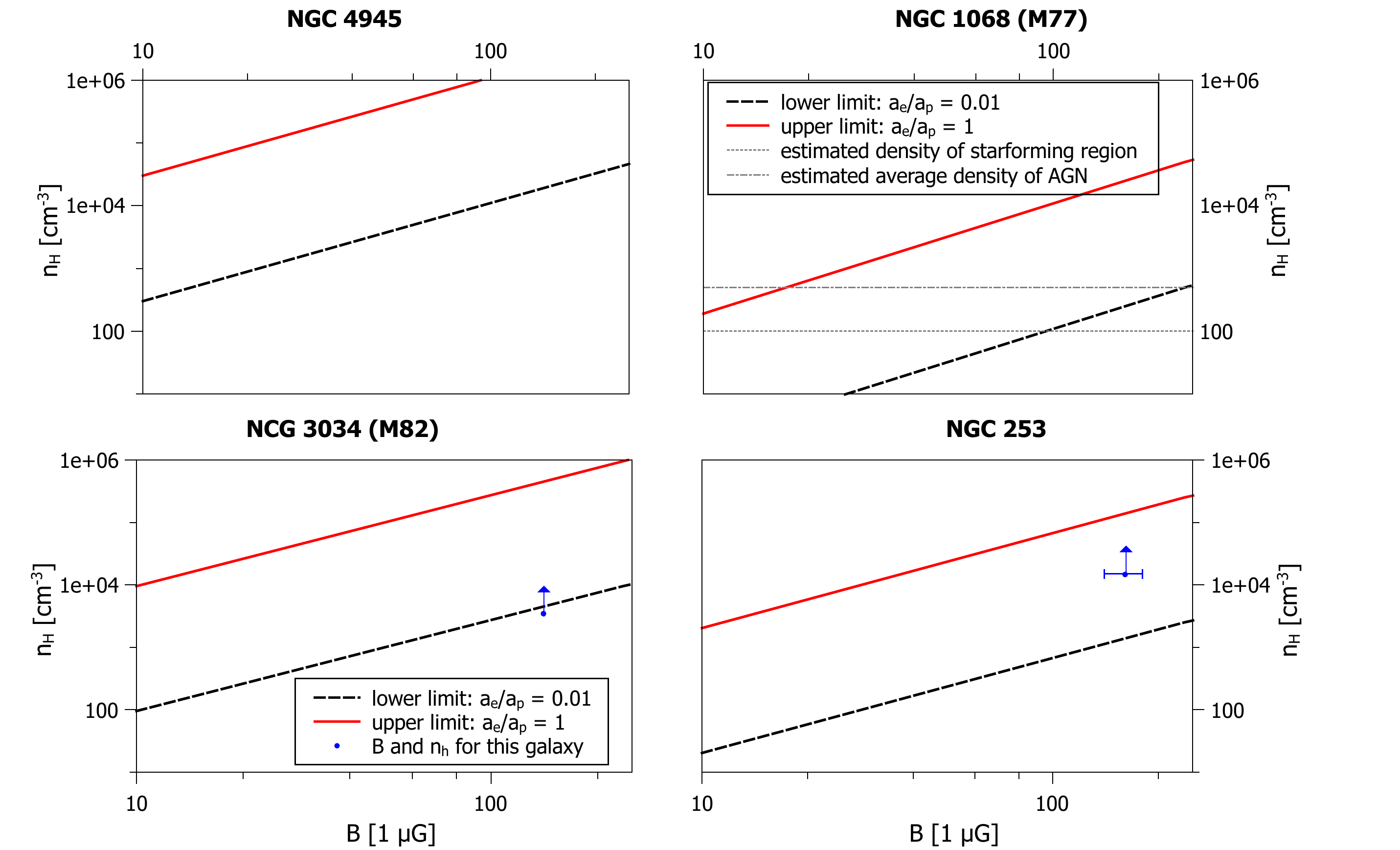}
\caption{Possible range of the parameters $B$ and $n_{\mathrm H}$ for the individual galaxies. Density estimates taken from: \cite{spinoglio2005, matsushita2010}.}
\label{ssc::params}
\end{center}
\end{figure}

\begin{figure}[htbp]
\centering
\includegraphics[width=0.4\textwidth]{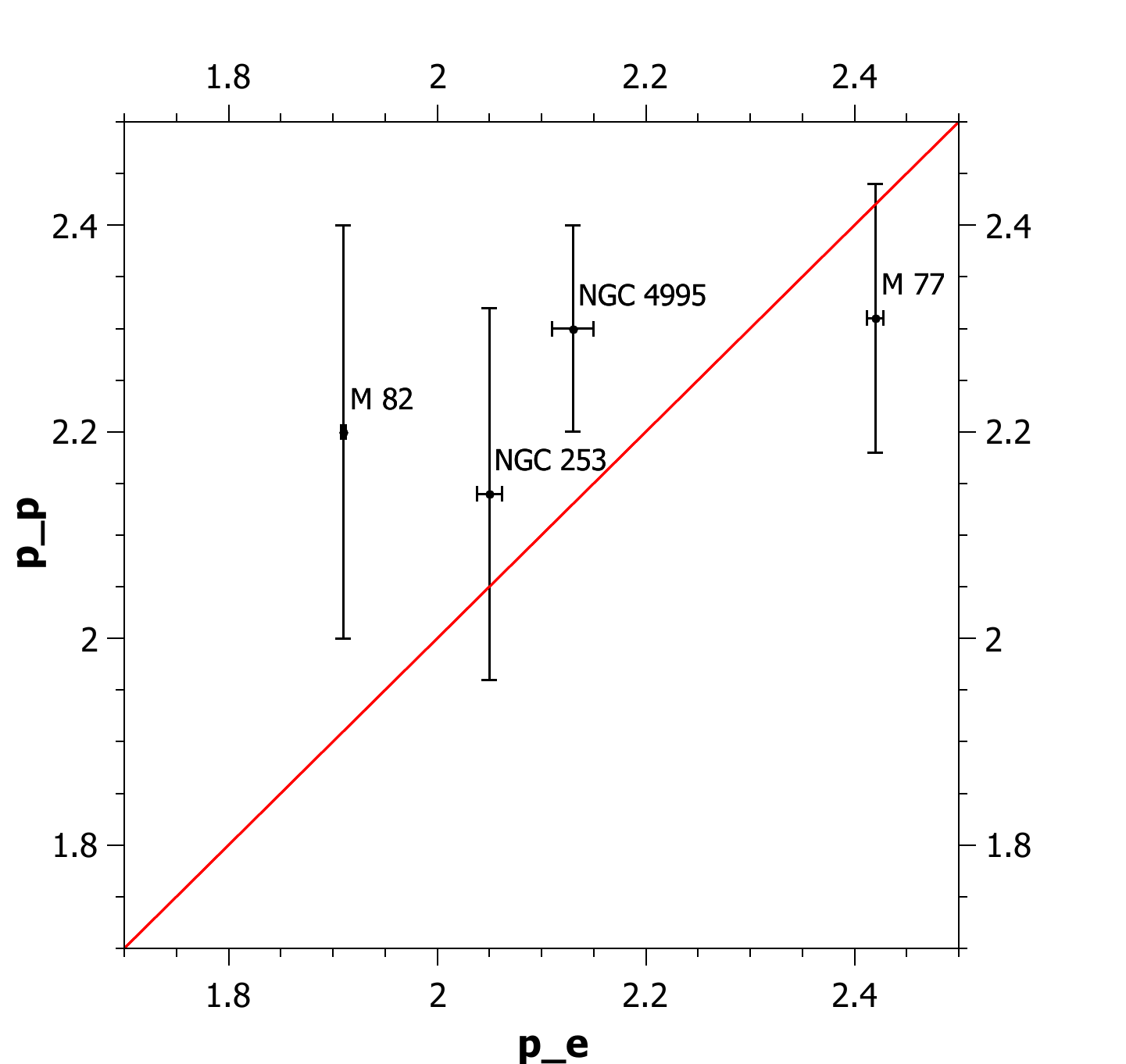}
\caption{Spectral index of the proton and electron distribution for each object as derived from gamma or radio data, respectively. The solid line represents $p_{\mathrm{p}}~=~p_{\mathrm{e}}$.}
\label{ssc::pp-plot}
\end{figure}

\begin{figure}[htbp]
\centering
\includegraphics[width=1.0\textwidth]{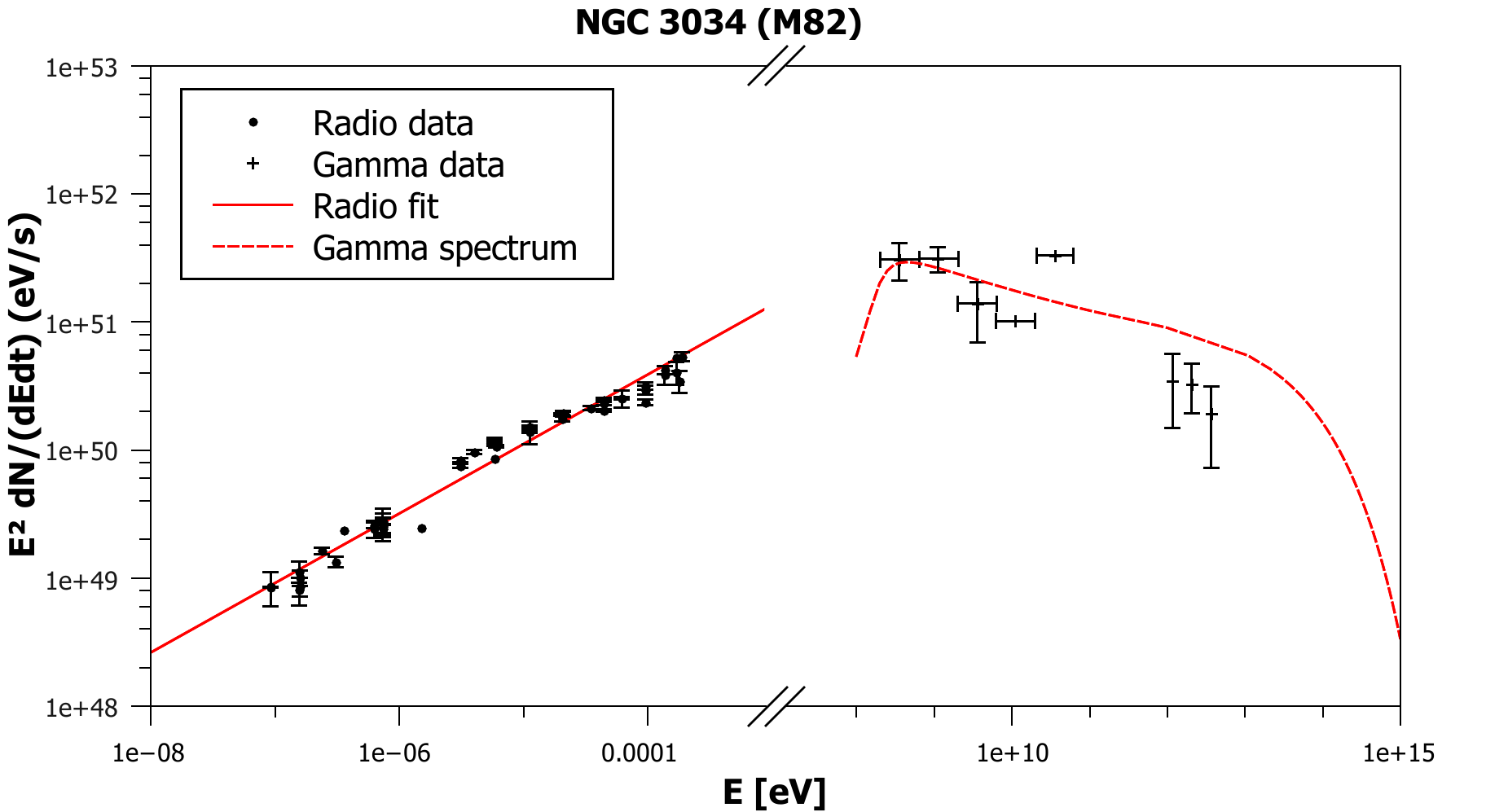}
\caption{Modeled partial SED of M82.}
\label{ssc::plot_SED_M82}
\end{figure}

Combining the spectra as derived from protons and electrons, respectively, allows to model the SED for each object examined. In Fig.\ \ref{ssc::plot_SED_M82}, this is shown for M82. The gamma ray luminosity $L_{\gamma}$ and the spectral index $\Gamma$ for NGC 4945, 1068 and 3043 given by \cite{lenain2010} and for NGC 253 from \cite{HESS2012} are used to calculate $A_{\mathrm{p}} = V N_{\mathrm {SNR}} \, a_{\mathrm{p}}$. Using radio data, fits to the radio spectra are made to find the radio synchrotron index $\delta$ and the normalization of the primary electron spectrum, $A_{\mathrm{e}}$, for a magnetic field of $B~=~1$~$\mu$G. From $\delta$, the spectral index $p$ of the underlying primary electron spectrum is derived. The corresponding values of the calculation are given in Table \ref{ssc::elecprotonpar}, where $n_{\mathrm H}~=~100$~cm$^{-3}$ was assumed for the calculation of $A_{\mathrm{p}}$. The quantity $A_{\mathrm{e}}$ depends on the magnetic field as $A_{\mathrm{e}}~\propto~{B^\delta}/{B}$, whereas $A_{\mathrm{p}}$ depends on the average hydrogen density $n_{\mathrm H}$ as $A_{\mathrm{e}}~\propto~{1}/{n_{\mathrm H}}$. These expressions can be combined using the inequality in eq.\ (3). This allows to determine a possible range of combinations of the magnetic field strength $B$ and the average hydrogen density $n_{\mathrm H}$. This is done in Fig.\ \ref{ssc::params}. As can be seen, for each galaxy the observed values for the average hydrogen density and the magnetic field strength lie within the theoretically allowed range, which is the area between the lines corresponding to ${a_{\mathrm{e}}}/{a_{\mathrm{p}}}=0.01$ (dashed black line) and ${a_{\mathrm{e}}}/{a_{\mathrm{p}}}= 1$ (solid red line). To test the assumption that the energy spectra of protons and electrons have the same shape, the electron and proton spectral indices of the individual galaxies have been calculated and are shown in Fig.\ \ref{ssc::pp-plot}.
\begin{table}[h]
\begin{center}
\begin{tabular}{|l|c|c|c|c|c|c|}
\hline
Source & $\Gamma$=$p_\mathrm{p}$ &  $A_{\mathrm{p}}
[\textrm{\small{1/erg}}]$ & $\delta$ & $p_\mathrm{e}$ & $A_{\mathrm{e}}
[\textrm{\small{1/erg}}]$ \\
\hline
NGC~1068 & $2.31\pm0.13$ &  1.45E+68  & $-0.710 \pm 0.004$ & $2.421 \pm 0.008$
&1.76E+69
\\
NGC~4945 & $2.3\pm0.1$ &  6.16E+67 & $-0.564 \pm 0.011$ & $2.129 \pm 0.022$ &
7.50E+66
\\
NGC~253 & $2.14\pm0.18$ & 1.00E+66 & $-0.525 \pm 0.006$ & $2.051 \pm 0.012$
&1.77E+66
\\
NGC~3034 & $2.2\pm0.2$ &  3.20E+66 & $-0.457 \pm 0.001$ & $1.914 \pm 0.002$
&9.60E+65
\\
\hline
\end{tabular}
\caption[Parameters $A_{\mathrm{p}}$ and $A_{\mathrm{e}}$ from the radio GeV correlation]{Parameters
$A_{\mathrm{p}}$ calculated from $\Gamma$-index and gamma ray luminosity, $A_{\mathrm{e}}$ calculated from a fit to the radio data.}
\end{center}
\label{ssc::elecprotonpar}
\end{table}
For a given average hydrogen density and an estimate of ${a_{\mathrm{e}}}/{a_{\mathrm{p}}}$, this approach could be used to determine the magnetic field strength, complementary to the methods presented in \cite{thompson2006} and \cite{heesen2011}. For a given average hydrogen density, magnetic field strength and a fixed ratio for ${a_{\mathrm{e}}}/{a_{\mathrm{p}}}$, this approach could be used to check whether the galaxy is calorimetric for both protons and electrons to the same extent. However, further improvement of the statistics is required to check the possible correlation of GeV gamma rays and radio emission from star-forming regions. Additionally, a more detailed theoretical examination has to be performed to establish this connection and gain new insights. The versatility of such a correlation strongly motivates further investigation in the future. 

\begin{acknowledgement}
We would like to thank B.\ Adebahr and R.J.\ Dettmar for helpful and inspiring discussions. We further acknowledge funding from the DFG, Forschergruppe "Instabilities, Turbulence and Transport in Cosmic Magnetic Fields" (FOR1048, Project BE 3714/5-1), the support from Mercator Stiftung, with a contributing grant within the MERCUR project An-2011-0075, from the Junges Kolleg (Nordrheinwestf\"alische Akademie der Wissenschaften und der K\"unste) and the support by the Research Department of Plasmas with Complex Interactions (Bochum).
\end{acknowledgement}

\end{document}